\documentstyle[a4,12pt]{article}
\begin{document}
\textwidth    170mm
\textheight   250 mm
\pagestyle{plain}
\hoffset 0 mm
\topmargin -1 cm
\newcommand{\be}{\begin{equation}}
\newcommand{\ee}{\end{equation}}
\newcommand{\bea}{\begin{eqnarray}}
\newcommand{\eea}{\end{eqnarray}}
\newcommand{\nn}{\nonumber}
\newcommand{\muh}{\hat\mu}
\newcommand{\dlr}{\stackrel{\leftrightarrow}{D} _\mu}
\newcommand{\vnew}{$V^{\rm{NEW}}$}
\newcommand{\vecp}{$\vec p$}
\newcommand{\dof}{{\rm d.o.f.}}
\newcommand{\prd}{Phys.Rev. \underline}
\newcommand{\pl}{Phys.Lett. \underline}
\newcommand{\prl}{Phys.Rev.Lett. \underline}
\newcommand{\np}{Nucl.Phys. \underline}
\newcommand{\vvp}{v_B\cdot v_D}
\newcommand{\dl}{\stackrel{\leftarrow}{D}}
\newcommand{\dr}{\stackrel{\rightarrow}{D}}
\newcommand{\mev}{{\rm MeV}}
\newcommand{\gev}{{\rm GeV}}
\newcommand{\calp}{{\cal P}}
\newcommand{\ra}{\rightarrow}
\newcommand{\Ra}{\Rightarrow}
\newcommand{\la}{{\cal L}}
\newcommand {\fmn}{F_{\mu\nu}F^{\mu\nu}}
\newcommand {\gmn}{G^a_{\mu\nu}G_a^{\mu\nu}}
\newcommand {\fuq}{\frac 1 4}
\newcommand {\wmn}{W^a_{\mu\nu}W_a^{\mu\nu}}
\newcommand {\bmn}{B_{\mu\nu}B^{\mu\nu}}
\def\dsl#1{\mathchoice
 {\dslaux\displaystyle{#1}} {\dslaux\textstyle{#1}} {\dslaux\scriptstyle{#1}}
 {\dslaux\scriptscriptstyle{#1}} }
\def\dslaux#1#2{\setbox0=\hbox{$#1{#2}$}
 \rlap{\hbox to \wd0{\hss$#1/$\hss}}\box0}
\let\sla=\dsl
\def\eqti{ {\,\,\lower.7ex\hbox{$=\atop t\rightarrow\infty$}\,\,}}
\def\mqi{ {\,\,\lower .8ex\hbox {$\longrightarrow
 \atop m_Q \rightarrow \infty$}\,\,}}
\newcommand{\linf}[1]  {\,\,\lower .8ex\hbox {$\longrightarrow
 \atop #1 \rightarrow \infty$}\,\,}
\def\ni{\noindent}
\def\nl{\newline}
 LPTHE-Orsay 94/54//
 hep-ph/9403341

\begin{center}
{\Large\bf B DECAYS, AN INTRODUCTORY SURVEY.}

\vskip 1.5cm
 A. Le Yaouanc, L Oliver, O.P\`ene, J.-C. Raynal\linebreak
LPTHE, F 91405 Orsay, France,\footnote {Laboratoire
associ\'e au Centre National de la Recherche Scientifique.}\linebreak
Presented by O. P\`ene.
\end{center}
\vskip 2 cm
\begin{abstract}

To start  the $b$-decay session
we briefly introduce and comment some important theoretical tools which are
currently used in $b$ physics. Heavy Quark Symmetry and its consequences for
heavy to heavy and heavy to light semi-leptonic decays, as well as for leptonic
decays, are briefly summarised. It is stressed that symmetry must be completed
with dynamical calculations.
A critical discussion of the nearest pole dominance (VMD) assumption is
performed.  Parton model and its higher twist corrections are discussed on the
example of lifetimes. Finally  non-leptonic decays are considered via the
example of the exclusive calculation of $\Delta \Gamma$ in  the $B_s-\overline
B_s$ system. The popular factorization assumption is discussed and seems to be
rather good.

   \end{abstract}

\section{What is so exciting about beauty?}
\label{sec-what}

This talk should be taken as a short introduction to the following ones on
beauty decay. Beauty physics has become one of the most active fields. It would
be difficult to quote one large experimental  device able to produce $b$'s
which is not making of the study of $b$ decays one of its priority. Several
thousand physicists are actively engaged in this field. What is the reason for
this growing enthousiasm? Let us try some answers.

\paragraph {Beauty is heavy,} while
charm is not so heavy and top will not produce hadrons. This implies several
theoretical niceties: Heavy Quark Symmetries, validity of parton
model, etc. It also implies that there are
many final states and hence a rich phenomenology.

\paragraph {Beauty is rather stable.}

 The coupling of the third generation to the other two turns out to be rather
small,
($O(\lambda^2)$) in Wolfenstein's parameterization. This lucky feature raises
serious hopes that CP violation may be measured in B decay.
In fact, a long standing intimacy has existed between beauty and CP violation,
since the third generation was postulated by
 Kobayashi and Maskawa because their
mechanism for CP violation needed it. But CP violation is out of the scope of
this talk and we shall leave it aside, after having reminded you that it is
presumably the main reason of the widely spread enthousiam for $B$ physics.

\paragraph{The worst known CKM parameters concern the third generation:
$V_{ub}$, $\sin \delta$, ..}
We need these parameters, for the sake of SM itself, and hopefully to learn
something unexpected beyond it.

\section{ Main Theoretical tools  for b decays.}

We will restrict ourselves to heavy-light hadrons: $Q\bar q$ mesons and
$Q q q'$ baryons where $Q$ represents any heavy quark and $q, q'$  any light
quark. Let us make a list of the main concepts, principles, rigourous methods,
models, etc. that are commonly used:
\begin{itemize}

\item { Heavy Quark Symmetry}
\item{  Duality and Parton Model}
\item{  Lattice QCD}
\item{  QCD Sum Rules}
\item{  Quark Models}
\item{  Analiticity, Vector Meson Dominance,...}
\item{  Factorization assumption for non-leptonic decays.}
\end{itemize}

We will try to say a few words on these items, except for lattice calculations
that will be considered later by Asmaa Abada \cite{asmaa} and QCD sum rules on
which Stephan Narison will give a review. For lack of time we will also skip
quark models although they are able to give a precious physical insight. Recent
analyses of the latter may be found in \cite{aleksanpsi}.

\section{ Heavy Quark Symmetry (HQS)}

It is strange that the HQS has only been understood
recently, although it is a direct consequence of QCD. It was anticipated
by some works in the late eighties and fully emphasized only four years
ago. It is impossible to quote but a few among the huge number of
publications it has triggered in such a short period: \cite{eic}-\cite{wis}.
Although you must all have heard of HQS, it is unavoidable to say some words
about it.

The basic idea may be explained through the ``atomic picture'':
Up to small corrections, the properties of an atom depend only on the
electric charge of the nucleus (the atomic number), not of its atomic mass.
This is
because the nucleus is so heavy vz the electron that it is practically at rest
in the
center of mass of the atom, and it acts as a static electric charge.
Mutatis mutandis, this happens when a heavy quark is bound with a light
quark. Up to $O(1/m_Q)$ corrections, the heavy quark acts as a triplet
static source of color.

In other words, let us write the heavy quark momentum as
$p^\mu= m_Q v^\mu + k^\mu$
where $m_Q$ is the heavy quark mass, $v_\mu$ is the hadron four velocity
(momentum divided by the mass), and $k_\mu$ is a momentum that represents
the effect of the wave function, and it is of the order of the QCD
scale, remaining constant when $m_Q\ra \infty$.

\be \frac {i m_Q \sla v + \sla k}{(m_Q v+k)^2-m_Q^2} \simeq \frac {i \sla v +1}
{2 v\cdot k}\ee
where the $\simeq$ symbol here means up to $O(1/m_Q)$ corrections.
   Flavor symmetry is obvious since no dependence on the heavy mass is left.
Let us now
assume many soft gluons emitted from the heavy line:

\bea &\bar u(s,v)  \frac {i \sla v +1}{2} it_a\gamma^\mu \frac{i \sla v +1}{2}
it_b\gamma^\nu \frac{i \sla v +1}{2}...= \nn \\
&\bar u(s,v) \frac {i \sla v +1}{2}
t_a t_b ... v^\mu v^\nu ...\nn\eea

No dependence on the heavy quark spin is left,
and we end up with a  $SU(2 N_F)_v$ symmetry where $N_F$ is number of heavy
flavors. One should notice however that the symmetry acts inside a sector
corresponding to one heavy hadron velocity. This is the meaning of the
index~$v$.

\section{ Scaling laws and symmetry relations in the heavy quark limit.}

We turn to the phenomenological consequences of HQS on leptonic decays, heavy
to heavy semi-leptonic decays and finally heavy to light semi-leptonic decays.
Although the second one is the best known, the other two are also very useful
for phenomenology. In all this section we will neglect anomalous dimensions.

\subsection {Scaling laws for letponic decay constants: $B, D\ra l \nu$}

Let us simply state the result \cite{eic}. The leptonic decay constant of a
heavy pseudoscalar meson $P$ composed of a heavy quark $Q$ and a light one
$\bar q$ scales like

\be F_P M_P^{1/2} \linf {m_Q}  \hbox{ contant} + O(\frac 1 {M_P})\ee
 and, of course $M_P\simeq m_Q$. The same is true for a vector meson with the
same constant.
But the symmetry does not tell us:

\ni How much is the constant?\newline
How large are the $1/M_P$ corrections? what is their sign?\newline

 For the answer we need to ask  Lattice QCD and QCD sum rules
\cite{asmaa},\cite{narison}. These two methods yield the same result: the
$1/m_Q$ are such as {\it to soften the $1/\sqrt{M_P}$ decrease of $F_P$} as a
function of $M_P$. In practice, the predictions for $F_B$ are not very
different from those for $F_D$ while scaling rules without corrections would
predict $F_B / F_D \simeq 0.6$.

\subsection {Scaling laws for Semi-leptonic decay $B\ra D l \nu$}
\label{sec-hh}

{
\bea <D(v')\vert V_\mu \vert B(v)> \mqi& \sqrt {M_B M_D}\,\,\xi(v\cdot
v') (v+v')_\mu\nn \\
<D^\ast(v')\vert V_\mu \vert B(v)>\mqi& \sqrt {M_B
M_D}\,\,\xi(v\cdot v')\epsilon_{\mu\nu\alpha\beta}\epsilon^{\ast \nu}
v'^\alpha v^\beta\nn \\
<D^\ast(v')\vert A_\mu \vert B(v)>\mqi& \sqrt {M_B
M_D}\,\,\xi(v\cdot v') (v\cdot v' + 1 -
\epsilon^\ast\cdot v v'_\mu)\label{iw} \eea}
where $\xi(v.v')$ is the celebrated Isgur-Wise function \cite{wis}, with
$\xi(1) = 1$ (corresponding to $v'=v$). But:

\ni $\xi(y)$  is unknown for $y\ne 1$\nl
 How large are the $)(1/m_Q)$ corrections?\nl

This scaling law has already proven to be very useful for phenomenology.
Neubert \cite{neubert} has proposed a direct use of the relations (\ref{iw})
and of the normalization $\xi(1)=1$ to measure  $\vert V_{cb}\vert$ from $B\ra
D^\ast l \nu$.   Using this method CLEO \cite{grady} obtains  $\vert
V_{cb}\vert= 0.037\pm 0.005\pm 0.004$.

\subsection { Scaling laws for Semi-leptonic decay $B\ra K(^\ast),\pi, \rho l
\nu$
}{ In the rest frame of the initial (heavy) meson \cite{sliw}:

$$\frac {<K(^\ast), \vec p \,\vert J_\mu \vert B>}{\sqrt{M_B}}\simeq\frac
{<K(^\ast), \vec p \,\vert J_\mu \vert D>}{\sqrt{M_D}}$$
if
$$\vert \vec p\vert, M_K, M_{K^\ast}\ll M_D, M_B$$

Notice that small $\vert \vec p\vert$ means $q^2 \sim q_{max}^2$. Once again,

\ni the symmetry provides a relation between matrix elements, but not
their values\nl
 neither the size and sign of the corrections.
\nl

Lattice QCD \cite{asmaa} gives a preliminary answer: for $A_1$ and $V$ the
corrections tends to soften the $M_P$ dependence of the dominant term
(similarly to the case of $F_P$), $A_2$ presents the opposite trend but with
large errors, and $f_+$ is close to the uncorrected scaling. A phenomenological
analysis \cite{aleksanpsi}  of $B\ra\Psi K^\ast$ tends to confirm the general
trend of a ``soft'' scaling which can be understood very simply in the quark
models \cite{aleksanpsi}, \cite{lopr}.

\section { $q^2$ dependence of form factors :  Beware of Vector Meson Dominance
!}

Little is known about the $q^2$ dependence of the form factors ($q^2$ is the
invariant squared mass of the final leptons). The scaling laws described in
section \ref{sec-hh} predict only the $q^2$ dependence of the {\it ratio}
between different heavy to heavy form factors. But this limited piece of
information is missing in the case of heavy to light form factors. There is a
wide-spread belief that the nearest pole dominance should not be a bad
approximation, i.e.

\be F(q^2) \propto \frac 1 {q^2 - M_{B'}^2}\ee
where we call $B'$ the lightest state which has the  quantum numbers exchanged
in the $t$ channel, for example $B^\ast, B^{\ast \ast}$, .... If we consider,
say, $B\ra \pi l \nu$, the physical region corresponds to $0\le q^2 \le
q^2_{max}= (M_B-M_\pi)^2 < M^2_{B'}$. Pole dominance is valid only in the
vicinity of the pole. It may be valid near $q^2_{max}$ when the pole is not too
far away, but it is certainly not valid in the whole range. Many other poles,
cuts, etc contribute. If lattice estimates \cite{asmaa} seem to favor form
factors that increase with $q^2$ near  $q^2_{max}$, QCD sum rules seem to
indicate no pole dominance for axial form factors\cite{ball}, and a
phenomenological analysis of $B\ra\Psi K^\ast$ \cite{aleksanpsi} also seems to
discard pole dominance in favor of a differentiated behaviour for the various
form factors. A weak binding relativistic quark model predicts $A_1$ to be
flatter than $f_+$\ \cite{aleksanpsi}, \cite{lopr}.

\section {  Duality and parton model,  life times.}

\subsection{Plain parton model.}

 Parton model assumes that
the total width $\Gamma(B\ra \hbox{everything}) \simeq \gamma(b\ra c q \bar q,
c l \bar \nu)$. It is based on the idea that the spectator quark plays no role
because the final
quarks are hard. Plain parton model then predicts

\centerline{$ \tau_{B^0}= \tau_{B^-} = \tau_{\Lambda_b}$}

as it would predict  $\tau_{D^0}= \tau_{D^+} = \tau_{\Lambda_c}$ if charm was
assumed to be heavy.\hfill\newline  The latter assumption is in  total
contradiction with present experimental values:
$$ \tau_{D^0}= 4.20\pm 0.08\, 10^{-13}s,\quad \tau_{D^+}= 10.66\pm 0.23\,
10^{-13}s,\quad \tau_{\Lambda_c}=1.91 {+0.15\atop-0.12}10^{-13}s $$
\noindent and $\tau_{\Xi_c}=3.20 {+0.9\atop-0.8}10^{-13}s$ \cite{simon}. We see
that  charm cannot be considered as heavy under this respect.

 What about beauty? The present situation \cite{hessing} is:

\be \frac {\tau_{B^-}}{\tau_{B^0}}=1.14\pm 0.15,\quad
\frac {\tau_{B_s}}{\tau_{B^0}}=1.11\pm 0.18,\quad
\frac {\tau_{\Lambda_b}}{\tau_{B^0}}=0.75\pm 0.12. \ee

\noindent This looks much better than for the charm, but
${\tau_{\Lambda_b}}/{\tau_{B^0}}$ is two sigmas away from the uncorrected
parton model prediction.

 \subsection {How to compute the corrections to parton model?}

A recent series of papers \cite{bigi}
 propose a  generalization of parton
model:

 \be\Gamma(H_Q\ra X ) \propto G_F^3 < H_Q \vert Im \hat T(Q\ra X\ra Q)\vert
H_Q>\ee
is expanded into matrix elements of the operators $v_\mu\overline
Q\gamma^\mu Q$,\hspace{0.3 cm} $\overline Q (D^2 -(v\cdot D)^2) Q$, \hspace{0.3
cm} $\overline Q
\sigma^{\mu\nu}G_{\mu\nu} Q$, \hspace{0.3 cm} $\overline Q\Gamma q \bar q
\Gamma Q$, etc.

\paragraph{The first operator,} $ \overline Q \dsl v Q$, of dimension three,
gives the parton model. Next, there are no $1/m_Q$ corrections.

\paragraph{The $O(1/m_Q^2)$ corrections} are generated by the  dimension five
operators.   The $\overline Q (D^2 -(v\cdot D)^2) Q$ corresponds to the Fermi
motion. The $\overline Q
\sigma^{\mu\nu}G_{\mu\nu} Q$ matrix element is known from hyperfine splitting
between meson masses. These dimension five operators do not split the meson
life-time degeneracy. To $O(1/m_Q^2)$

\be \tau_{\Lambda_b}\ne \tau_{B^0}= \tau_{B^+}
\Ra  \frac {\tau_{\Lambda_b}-\tau_{B^0}}{\tau_{\Lambda_c}-\tau_{D^0}}
\simeq \frac {m^2_c}{m^2_b}.\ee

It is gratifying that experiment seems to confirm that $\vert
{\tau_{\Lambda_b}-\tau_{B^0}}\vert > \vert{\tau_{B^+}-\tau_{B^0}}\vert$.
However it is still unclear if this model can account for such a small
${\tau_{\Lambda_b}}/{\tau_{B^0}}$ ratio as 0.75.

\paragraph{The most interesting $O(1/m_Q^3)$ operator} is $\overline Q\Gamma q
\bar q \Gamma Q$ since it produces the first
non-spectator effect (the light quark fields can act on the spectator quarks).
It results:
\be \frac {\tau_{B^+}-\tau_{B^0}}{\tau_{D^+}-\tau_{D^0}}
\simeq \frac {m^3_c}{m^3_b}\ee

Finally, this approach is new and still under discussion, however it has
several nice features and certainly deserves further work.

\section { Non-leptonic decays and factorization assumption}

\subsection{Some general remarks.}

Non leptonic decays represent the dominant decay channels of the $B$'s. But
there exists no exact method to deal with them. One usually resorts to the
factorization assumption which
allows to express two body non leptonic decays as a product of a semi-leptonic
amplitude and a purely leptonic one. It neglects the soft-gluon exchange
between the two parts of the diagram. It is only
valid when $N_c \ra \infty$, but in $B$ physics it often appears to be a
reasonable approximation with phenomenological coefficients, $a_1, a_2$,
multiplying both relevant four quark operators. CLEO \cite{henderson} has
carefully checked factorization for $N_c$ dominant decays, and the success of
factorization seems to extend to their fit of the $a_2/a_1$ ratio which is
$O(1/N_c)$.

There is an overwhelming number of studies of non-leptonic decays. We will give
a recent example \cite{aleksan}  noticeable, in the realm of non-leptonic
decays, by the fact that a calculation of phenomenological interst turns out to
be unexpectedly under control.

\subsection {An example: exclusive computation of $\Delta \Gamma$ in
$B_s-\overline B_s$}

The $B_s-\overline B_s$ system has
$\Delta M/\Gamma \gg 1$, which makes it difficult to measure \cite{tarem} the
too fast oscillations $\sim \sin (\Delta M \,t)$.
 What about $\Delta \Gamma$?
Is $\Delta \Gamma/ \Gamma$ large enough to allow it to be seen and
used?\footnote{Remember how useful has been, in the $K-\overline K$ system, the
fact that $\Gamma(K_S)\gg \Gamma(K_L)$ and hence the fact that only $K_L$'s fly
a few meters before decaying.}

Inclusive parton model calculations \cite{hagelin} give: $\Delta \Gamma/
\Gamma\simeq 0.20$. The question is, what are the effects of the lightest
exclusive channels?  Should they be added to parton model contribution? Are
they dual to it?
The analysis \cite{aleksan} leads to the following conclusions:

\begin{itemize}
\itemsep=-1 mm
\item The dominant decay channels are $D_s \overline D_s,\,D_s^{\ast} \overline
D_s,\,D_s \overline {D_s^{\ast}},\,D_s^{\ast} \overline {D_s^{\ast}}$.
\item These channel contribute mainly with the same sign to $\Delta \Gamma$.
Their contribution is known in a model independent way from experimental
\cite{PDG} $B_d\ra D_s^{(\ast)} \overline{D^{(\ast)}}$ branching ratios.
\item Factorization is a reasonable assumption in this case.
\item Exclusive and inclusive calculations are dual: they are not to be added,
they should approximately agree,  and they do !
 \item $\Delta \Gamma/ \Gamma\simeq 0.15 $
\end{itemize}

This is maybe the largest lifetime difference among the $B$'s, and it is
measurable \cite{cen} !

\section{ Conclusions}

\begin{itemize}
\item {} Heavy Quark Symetry is a simple and powerful consequence of QCD. But
it needs to be completed by dynamical computations of universal constants,
universal functions, and $1/m_Q$ corrections.
\item {} There are several instances where the $1/m_Q$ corrections tend to
soften the dependence on the heavy mass indicated by the dominant term.
 \item {} The corrections to parton model for inclusive processes may be
understood via a systematic expansion in higher dimension operators.
\item {} Lattice QCD predicts leptonic decays and semi-leptonic ones (for not
too large momenta).

\item {} The $q^2$ dependence of the form factors is still largely unknown.
Nearest pole dominance has no theoretical grounding and there are indications
of other behaviours. QCD sum rules may help in this problem.

\item {} Non-leptonic decays are tractable, using the factorization assumption
which seems not too bad. Improvement needs an understanding of the corrections
to it, and of duality.

\noindent {\bf Acknowledgments :} \par
 I would like to thank the organisers of the rencontres, with a particular
thought for Jo\"elle Raguideau who was wounded the last day. This work was
supported in part by the CEC Science Project SC1-CT91-0729.\par

\end{itemize}


\begin{thebibliography}{999}
\bibitem{asmaa}As. Abada, these proceedings.
\bibitem{narison}S. Narison, these proceedings.
\bibitem{aleksanpsi}R. Aleksan, A. Le Yaouanc, L. Oliver, O. P\`ene, J.-C.
Raynal, LPTHE-Orsay 94-15 (1994).
\bibitem{lopr} A. Le Yaouanc et al., Gif lectures 1991 (electroweak properties
of heavy quarks), tome 1, p89; A. Le Yaouanc and O. P\`ene, Third workshop on
the Tau-Charm Factory, 1-6 june 1993, Marbella, Spain, Bulletin Board: hep-ph
9309230;
R. Aleksan et al., Beauty 94 workshop, april 1994, Mont Saint Michel, France
(to be published in Nucl. Inst. and meth. A) Bulletin Board: hep-ph 9406334.
\bibitem{eic} M.B.Voloshin and M.A.Shifman,
Sov.J.Nucl.Phys. \underline{45} (1987) 292;
Sov.J.Nucl.Phys. \underline{47} (1988) 511\\
 H.D.Politzer and M.Wise,
Phys.Lett. \underline{B206} (1988) 681;
Phys.Lett. \underline{B208} (1988) 504;\\
 E.Eichten,
Nucl.Phys. \underline{B(Proc.Suppl.)4} (1988) 170;
\bibitem{sliw}  N. Isgur and M.B. Wise, \prd{D42} (1990) 2388.
\bibitem{wis}
 H.Georgi, HUTP-91-A014;\\
J.D.Bjorken, SLAC-PUB-5362, SLAC-PUB-5389;\\
N.Isgur and M.Wise,
Phys.Lett. \underline{B232} (1990) 113, Phys.Lett.\underline{B237} (1990) 527,
Phys.Rev.Lett. \underline{66} (1991) 1130;\\
H.Georgi and F.Uchiyama,
Phys.Lett.\underline{B238} (1990) 395;\\
H.Georgi and M.Wise,
Phys.Lett. \underline{B243} (1990) 279.
\bibitem{neubert} M. Neubert Phys.Lett.\underline{B264} (1991) 455.
\bibitem{grady} C. O'Grady, these proceedings.
\bibitem{ball}P. Ball, Phys. Rev. \underline{D48} (1993) 3190.
\bibitem{simon} A. Simon, these proceedings.
\bibitem{hessing} T. hessing, proceedings of the XIX Rencontre de Moriond '94
Electroweak Interactions and Unified Theories, Editions Fronti\`eres.
\bibitem{bigi}  cf. for instance I.I. Bigi CERN-TH-7050/93,  Invited talk at
Advanced Study Conference on Heavy Flavors, Pavia, Italy, 3-7,Sep 1993, and
references therein on the work of M. Shifman, M. Voloshin, B. Blok, N.
Uraltsev, A. Vainshtein and I. Bigi.
\bibitem{henderson} S. henderson, these proceedings.
\bibitem{aleksan} R. Aleksan, A. Le Yaouanc, L. Oliver, O. P\`ene, J.-C.
Raynal, Phys.Lett.\underline{B316} (1993) 567, \underline{B317} (1993) 173.
\bibitem{tarem} S. Tarem, these proceedings.
\bibitem{hagelin} S. Hagelin Nucl. Phys. \underline{B193} (1981) 123.
\bibitem{PDG} Particle Data Group, Phys. Rev \underline{D45} (1992) Part II.
\bibitem{cen} Y. Cen, these proceedings.
\end{thebibliography}
\end{document}